\documentclass[fp,twocolumn]{jpsj3}
\usepackage{txfonts}
\usepackage{color}
\usepackage{graphicx}
\usepackage[export]{adjustbox}

\title{$d$-Density Wave (DDW) Scenario Description of the New Hidden Charge Order in Cuprates}

\author{Imam Makhfudz\thanks{imakhfudz@gmail.com}}
\inst{Laboratoire de Physique Th\'{e}orique--IRSAMC, CNRS and Universit\'{e} de Toulouse, UPS, F-31062 Toulouse, France} 

\abst{
In this paper, we show that the theory of high $T_c$ superconductivity based on a microscopic model with $d$-density wave (DDW) scenario
in the pseudogap phase is able to reproduce some of the most important features of the recent experimentally discovered hidden charge order in several families of Cuprates. 
In particular, by computing and comparing energies of charge orders 
of different modulation directions derived from a full microscopic theory with $d$-density wave scenario, the axial charge order $\phi_{X(Y)}$ with wavevector $\mathbf{Q}=(Q_0,0)((0,Q_0))$ 
is shown to be unambiguously energetically more favorable over the diagonal charge order $\phi_{X\pm Y}$ with wavevector $\mathbf{Q}=(Q_0,\pm Q_0)$ at least in commensurate limit, 
to be expected also to hold even to more general incommensurate case, in agreement with experiment. The two types of axial charge order $\phi_X$ and $\phi_Y$ are degenerate by symmetry.
We find that within the superconducting background, biaxial (checkerboard) charge order is energetically more favorable than uniaxial (stripe) charge order, and therefore checkerboard axial charge order 
should be the one observed in experiments, assuming a single domain of charge ordered state on each CuO$_2$ plane.
}

\newcommand{\comment}[1]{}

\begin{document}
\maketitle

\section{Introduction}

The mechanism of high $T_c$ superconductivity in cuprates is one of the biggest unsolved problems in condensed matter physics and remains not fully resolved and understood ever since its discovery in 1986 \cite{Cuprates86} with hundreds of thousands of papers published on the subject.
Most of theories of Cuprates rest on the basic assumption that the parent compound can be described by Mott insulator \cite{Anderson}\cite{LeeNagaWen}, which is inherently
strongly correlated electron system which cannot be described by Landau Fermi liquid theory of weakly interacting electrons. A totally opposite point of view sees cuprate superconductivity just as an instability
away from Fermi liquid and should therefore be describable in terms of interacting electrons as interaction is adiabatically turned on, on top of weakly interacting Fermi liquid of metals \cite{LaughlinCritique}. 

Pseudogap phase remains one of the most mysterious parts in the cuprate phase diagram. It is characterized by a gap in the electron spectral function but without true long range coherence that characterizes superconducting order. 
Available theories of pseudogap phase propose that it be described by some type of symmety breaking order. The experimental results
on this pseudogap regime that accumulate over the years support this idea to some extent. Related to this, pseudogap phase has also been hypothesized to be a time reversal symmetry-broken state \cite{TRSBpseudogap} and several theories have been proposed 
to explain the phenomenon, in the form of loop current order \cite{Varma} and $d$-density wave (DDW) order \cite{NayakDDW,C-L-M-N,Tewari}. In view of the latter, it has been argued that DDW is a natural description of pseudogap phase where the $d$-wave symmetry of the 
symmetry breaking order in pseudogap phase evolves into $d$-wave symmetry of superconducting state (DSC) on lowering the temperatures \cite{Laughlin}.

After several years of pacificity, the field of high $T_c$ in cuprates heats up again with recent discovery of hidden ordered state in the underdoped regime of the phase diagram, within pseudogap phase \cite{Hoffman,Vershinin,Silva,Fujita,Ghiringelli,JChang,AchkarPRL1,AchkarPRL2,CominScience,rcomin,WuNat,WuNatComms,LeBoeuf}. 
In short, recent experiments observe what has been hypothesized as hidden charge order in the underdoped regime of several families of cuprates. This hypothesis was deduced from indirect signatures
of such charge order detected with various techniques including scanning tunneling microscopy \cite{Hoffman,Vershinin,Silva,Fujita}, resonant x-ray scattering \cite{Silva,Ghiringelli,JChang,AchkarPRL1,AchkarPRL2,CominScience,rcomin}, 
nuclear magnetic resonance \cite{WuNat,WuNatComms}, ultrasound study \cite{LeBoeuf}\cite{Shekhter}, and other methods. 
Motivated by this experimental discovery, various theoretical proposals have been put forward to explain the presence of such order \cite{sachdevlaplaca,EfetovNature,chubukov,palee,tsvelikchubu}.

One of the most intensively studied questions in this regard concerns the character of the charge order, especially the intra and inter unit cell structures of the order parameter.
The inter unit cell character concerns the ordering wavevector $\mathbf{Q}$ of the charge order. Different microscopic (or semi microscopic) theories so far lead to varying conclusions. 
Spin-fermion model \cite{chubukovSF} leads to axial charge order $\phi_X,\phi_Y$ with $\mathbf{Q}$ along $x$ or $y$ axis respectively \cite{EfetovNature,chubukov}.
In $t-J$ model, as first discussed in \cite{metlitsachdev} within linear hot spot approximation, the emergent $SU(2)$ particle-hole symmetry and Hartree-Fock calculation 
at quadratic level \cite{sachdevlaplaca,SauSachdev} put $d$-wave SC as the leading instability, followed by diagonal CDW $\phi_{X\pm Y}$ and then by axial CDW $\phi_X,\phi_Y$ as 
the next leading instabilities. 
In this paper, we use a full microscopic model of cuprates with DDW scenario and show, using variational calculation, 
that the axial charge order is exclusively energetically favorable and that the axial charge order must be biaxial (checkerboard) in character. 

\section{Model}

We consider a microscopic model given by the following Hamiltonian \cite{Laughlin} 

 \begin{equation}\label{Laughlinmodel}
H=H_0+\Delta H  
 \end{equation}
\begin{equation}
 H_0=-t\sum^{2N}_{\langle ij\rangle\sigma} c^{\dag}_{i,\sigma}c_{j\sigma}+h.c.+t'\sum^{2N}_{\langle\langle ij\rangle\rangle\sigma} c^{\dag}_{i,\sigma}c_{j\sigma}+h.c.
-\mu\sum_ic^{\dag}_{i\sigma}c_{i\sigma}.
\end{equation}
\[
\Delta H=U\sum^{N}_i n_{i\uparrow}n_{i\downarrow}+\frac{J}{2}\sum^{2N}_{\langle ij\rangle\sigma\sigma'}(c^{\dag}_{i\sigma}c^{\dag}_{j\sigma'}c_{j\sigma}c_{i\sigma'}-\frac{1}{2}c^{\dag}_{i\sigma}c^{\dag}_{j\sigma'}c_{j\sigma'}c_{i\sigma})
\]
\[
+V_t\sum^{2N}_{\langle ij\rangle\sigma\sigma'}(c^{\dag}_{i,\sigma}c_{j\sigma}+h.c.)(n_{i\sigma}+n_{j\sigma'}-\frac{1}{2})
\]
\begin{equation}
+V_n\sum^{2N}_{\langle ij\rangle\sigma\sigma'}c^{\dag}_{i,\sigma}c^{\dag}_{j,\sigma'}c_{j\sigma'}c_{i\sigma}
+V_c\sum^{2N}_{\langle ij\rangle}(c^{\dag}_{i,\uparrow}c^{\dag}_{i,\downarrow}c_{j\downarrow}c_{j\uparrow}+h.c.)
\end{equation}
The noninteracting Hamiltonian $H_0$ contains kinetic hopping up to second nearest neighbor and chemical potential terms. The interaction Hamiltonian $\Delta H$ consists of onsite repulsion $U$, antiferromagnetic spin exchange $J$, hopping renormalizing interaction $V_t$,
nearest-neighbor repulsion $V_n$, and Cooper pair hopping term $V_c$. We will perform mean field theory on this model Eq. (\ref{Laughlinmodel}) with the DDW scenario and show that axial charge order is exclusively preferred over diagonal charge order, in complete agreement 
with the experimental observations so far.     

\begin{figure}
  \centering
         \includegraphics[scale=0.275]{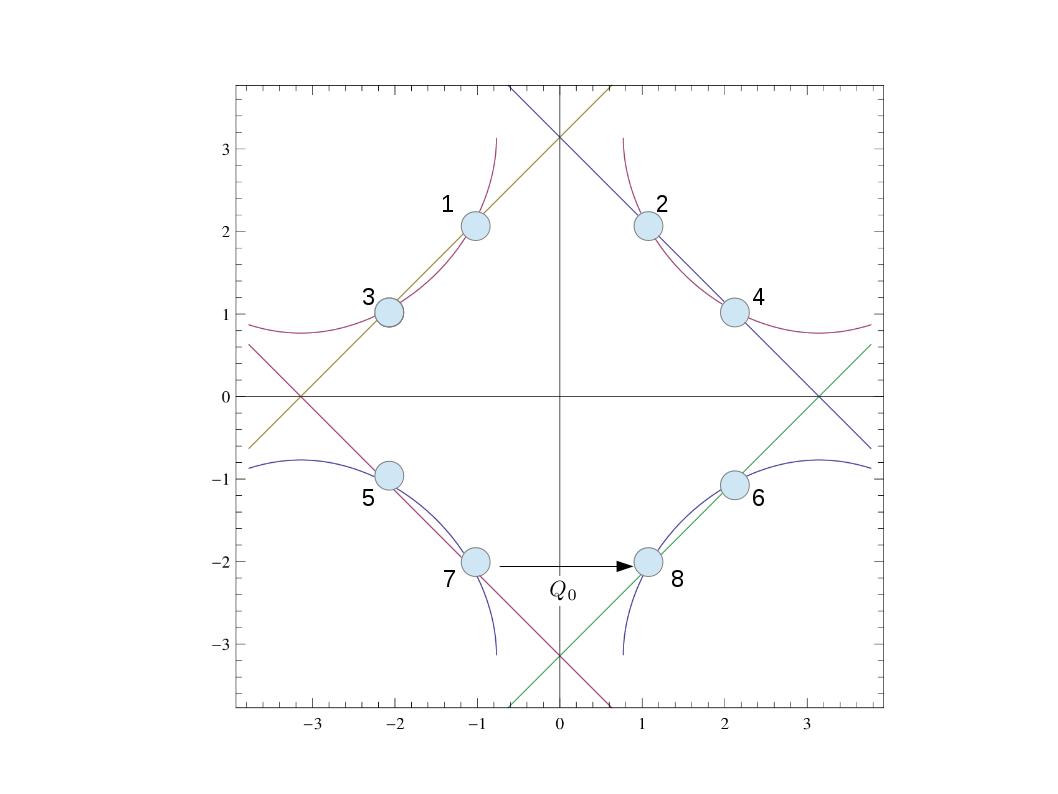}
  \caption{The Fermi surface and the hot spots (small circles) with $t=1.0,t'=0.3,\mu=-0.3$ and the incommensurate nesting wavevector $Q_0\approx 2.094=0.666\pi$ 
corresponding to periodicity $\lambda=3 a$, a reasonably realistic situation. All energy quantities used in this paper are given in eV (throughout this paper, we set the lattice spacing $a=1$).}
\label{fig:FSandHS}
\end{figure}

The noninteracting kinetic hopping Hamiltonian $H_0$ gives the usual tight binding dispersion $\epsilon_k=-2t(\cos k_x+\cos k_y)+4t'\cos k_x\cos k_y-\mu$ where $t,t'>0$ and normally $t'\ll t$. 
The nearest neighbor hopping produces square Fermi surface while the next nearest neighbor produces the curvature. 
The location of Fermi surface of free noninteracting fermions determined only by kinetic hopping term is given by $\epsilon_k=-2t(\cos k_x+\cos k_y)+4t'\cos k_x\cos k_y-\mu=0$
for the model with nearest neighbor and next nearest neighbor hoppings. The particle-hole scattering process in the density wave channel is mediated most importantly by 
antiferromagnetic exchange where only fermions near hot spots : points on the Fermi surface connected by ordering density wavevector (the 'nesting wavevector') contribute effectively to the scattering process.
To be consistent, we will also include spin antiferromagnetism order (SAF) into the full theory, which mediates such scattering process at appropriate wavevector.
The location of hot spots can be determined geometrically as given by the crossing points of the antiferromagnetic reciprocal lattice and Fermi surface. 
An example of Fermi surface with hot spots is shown in Fig.~\ref{fig:FSandHS} 
Experimentally, charge ordering is found to have periodicity with normally larger than two (around $3\sim 4$) multiple of unit lattice spacing \cite{Hoffman}\cite{Ghiringelli}\cite{JChang}\cite{CominScience}. 
This corresponds to incommensurate charge order. 
Its simplest description proposes that it be modeled by fermion scattering between nearby hot spots (e.g. hot spots 1 and 2 in Fig.\ref{fig:FSandHS}). The simpler case with commensurate charge order,
where $Q_0=\pi$, can be described by fermion scattering between hot spots on commensurate Fermi surface. In this case, the ordering wavevector of the charge order is the same as that of 
spin antiferromagnetism $\mathbf{Q}=(\pi,\pi)$. In the following sections, we will consider this comensurate limit and from that, make conjectures about the realistic incommensurate situation 
as seen in experiments.

\section{Variational Mean-Field Calculation}

Considering the commensurate limit, we decompose the interaction Hamiltonian $\Delta H$ into DSC, DDW, and SAF channels by deriving reduced Hamiltonians \cite{Nayak-Pivovarov}

\begin{equation}\label{redHDSC}
  H_{DSC}=-g_{DSC}\int_{k,k'}f_kf_{k'}c^{\dag}_{k,\uparrow}c^{\dag}_{-k,\downarrow}c_{k',\uparrow}c_{-k',\downarrow}
\end{equation}
\begin{equation}\label{redHDDW}
 H_{DDW}=-g_{DDW}\int_{k,k'}f_kf_{k'}c^{\dag}_{k+Q,\sigma}c_{k,\sigma}c^{\dag}_{k',\sigma'}c_{k'+Q,\sigma'}
\end{equation}
\begin{equation}\label{redHSAF}
 H_{SAF}=-g_{SAF}\int_{k,k'}c^{\dag}_{k+Q,\sigma}c_{k,\sigma}c^{\dag}_{k',\sigma'}c_{k'+Q,\sigma'}
\end{equation}
where $f_k=\cos k_x-\cos k_y$, giving the $d$-wave symmetry for both DDW and DSC, and $\int_k=\int d^2k/(2\pi)^2$. We simply have to determine what $g_{DSC}$, $g_{DDW}$, and $g_{SAF}$ are from Eq. (\ref{Laughlinmodel}). 
Naive decomposition of microscopic Hamiltonian Eq. (\ref{Laughlinmodel}) into different channels shows that not only $d$-density wave, $d$-wave superconducting, and spin antiferromagnetism states 
as given in Eqs. (\ref{redHDSC}),(\ref{redHDDW}),(\ref{redHSAF}) are present, but density wave and superconducting state 
of other symmetries also emerge naturally. In this case, we find $s$-wave density wave order (which is nothing but the conventional charge density wave (CDW)) and 
$s$-wave superconducting state and some states of other symmetries. 
This is inline with the fact that square lattice space group 
have several different irreducible representations (including the $d$-wave symmetry) each with its own basis functions, which however are found to have small even though finite weight for order parameters with other than $d$-wave symmetry \cite{sachdevlaplaca}.

The overall state is still heavily dominated by $d$-wave symmetry, both in particle-hole and particle-particle channels. 
we henceforth assume the $d$-density wave and $d$-wave superconducting states to be the dominant instabilities in pseudogap phase of cuprates, in addition to spin antiferromagnetism,
which is especially strong at lower doping level all the way to zero doping. 
This crucial point is in tune also with the physical argument that the hidden order in the pseudogap phase of cuprates 
should necessarily have $d$-wave symmetry because it is a crystal of Cooper pairs and is therefore continuation of the $d$-wave superconducting state at lower temperatures \cite{Laughlin}. 
We will show posteriori that this physical consideration is able to reproduce what has been observed in experiments on hidden charge order.

We obtain for the coupling constants in the reduced Hamiltonians Eqs. (\ref{redHDSC}),(\ref{redHDDW}),(\ref{redHSAF}) from Eq. (\ref{Laughlinmodel}) the following result.
\begin{equation}\label{gcouplings}
 g_{DSC}=6J-8V_n, g_{DDW}=6J+8V_n+8V_c,g_{SAF}=2(U+2J)    
\end{equation}
We see that the onsite $U$ term does not contribute to either DDW or DSC channel but contributes to SAF channel, as naively expected; strong onsite repulsion would prefer antiferromagnet order. 
The antiferromagnetic spin exchange coupling $J$ contributes to all channels
and reflects its dominant effect in dictating the energetics of the system and thus justifies our previous consideration regarding its role in determining the location of hot spots. 
The nearest neighbor repulsion 
with coupling constant $V_n$ manifests its effect in both $d$-wave channels. The $s$-wave Cooper pair hopping term with coupling $V_c$ in Eq. (\ref{Laughlinmodel}) contributes to the DDW channel. 
From Eqs. (\ref{redHDSC}),(\ref{redHDDW}),(\ref{redHSAF}), the $DDW$, $DSC$, and $SAF$ states are energetically favorable and may exist only if $g_{DSC},g_{DDW},g_{SAF}>0$. 
From Eq. (\ref{gcouplings}), with $U,J,V_t,V_n,V_c>0$ in Eq. (\ref{Laughlinmodel}),
we see that SAF is energetically favorable for large positive on-site repulsion $U$ and antiferromagnetic exchange $J$, DDW is always energetically favorable as we have $V_n,V_c>0$ and is bound to exist 
whereas DSC requires $J>J_c=4V_n/3$ to exist. It was found that these microscopic parameters do not depend on doping level
and in particular, fit to experiments suggests that $V_t=0$ \cite{Laughlin}, whereas the $U,J,V_c$ are all positive. 
These altogether secure the existence of DDW and DSC in the pseudogap regime of cuprates' phase diagram in addition to the spin antiferromagnetism SAF. 

Performing Hubbard-Stratonovich transformation, we write

\[
 Z=\int \mathcal{D}c^{\dag}\int \mathcal{D}c e^{-S[c^{\dag},c]}
\]
\begin{equation}
=\int \mathcal{D}c^{\dag}\int \mathcal{D}c \int \mathcal{D} \Delta \int \mathcal{D} \phi \int \mathcal{D} M e^{-S[c^{\dag},c,\Delta,\phi,M]}
\end{equation}
where, using finite temperature $T=1/\beta$ (imaginary) time-independent Euclidean space-time formalism, we have
\[
S_{[c^{\dag},c,\Delta,\phi,M]}=\int^{\beta}_0 d\tau H_{[c^{\dag},c,\Delta,\phi,M]}
\]
\[
=\int^{\beta}_0 d\tau[\sum_{\sigma}\int_k \epsilon_k c^{\dag}_{k\sigma}c_{k\sigma}+\frac{|\Delta|^2}{g_{DSC}}+\frac{|\phi|^2}{g_{DDW}}+\frac{|M|^2}{g_{SAF}}
\]
\[
+(\Delta^*\int_k f_kc^{\dag}_{k\uparrow}c^{\dag}_{-k\downarrow}+\phi^*\sum_{\sigma}\int_k f_kc^{\dag}_{k+Q\sigma}c_{k\sigma} 
\]
\begin{equation}\label{HSeffaction}
+ M^*\sum_{\sigma}\int_k c^{\dag}_{k+Q\sigma}c_{k\sigma} +h.c.)]
\end{equation}
where the DSC, DDW, and SAF order parameter fields are given by 

\begin{equation}\label{Deltadefinition}
\Delta=g_{DSC}\langle c^{\dag}_{i\sigma}c^{\dag}_{j,-\sigma}\rangle_{DSC}=g_{DSC}\int_k f_k \langle c^{\dag}_{k\uparrow}c^{\dag}_{-k\downarrow} \rangle
\end{equation}

\begin{equation}\label{phidefinition}
 \phi=g_{DDW}\langle c^{\dag}_{i\sigma}c_{j\sigma}\rangle_{DDW}=g_{DDW}\int_k f_k\langle c^{\dag}_{k+Q,\sigma}c_{k\sigma}\rangle
\end{equation}

\begin{equation}\label{Mdefinition}
 M=g_{SAF}\langle c^{\dag}_{i\sigma}c_{i\sigma}\rangle=g_{SAF}\int_k \langle c^{\dag}_{k+Q,\sigma}c_{k\sigma}\rangle
\end{equation}
It is to be noted that both the DDW and DSC orders considered in this theory are bond order type in real space; that is, they represent order parameters defined on Cu-O bond rather than at the Cu site \cite{C-L-M-N}\cite{Laughlin}. 
The (commensurate) DDW state breaks translational symmetry by one lattice spacing into two lattice spacings (i.e. doubles the unit cell) and breaks time reversal symmetry since it is represented by arrow of current on the bond (which corresponds to 
complex value for DDW order parameter $\phi$) and 4-fold rotational symmetry \cite{C-L-M-N}\cite{GScorrection}. 
DSC state is also defined on the bond; it breaks 4-fold rotational ($x$ vs. $y$) symmetry but still preserves translational symmetry by one lattice spacing and (with $d_{x^2-y^2}$ symmetry) the time reversal symmetry. 
The spin antiferromagnetism, on the other hand, is defined on the site by the staggered magnetization of the antiferromagnet N\'{e}el state, which therefore breaks translational symmetry by
enlarging the unit cell into two sites per unit cell and corresponds to ordering wavevector $\mathbf{Q}=(\pi,\pi)$.

We will consider and compare the energetics of four different states; (i) State with DSC, one diagonal DDW (with either $\mathrm{DDW}_{d+}$ with $\mathbf{Q}_{d+}=(\pi,\pi)$ or 
$\mathrm{DDW}_{d-}$ with $\mathbf{Q}_{d-}=(-\pi,\pi)$), and SAF (ii) State with DSC, two diagonal DDW 's with both $\mathrm{DDW}_{d+}$ with $\mathbf{Q}_{d+}=(\pi,\pi)$ and 
$\mathrm{DDW}_{d-}$ with $\mathbf{Q}_{d-}=(-\pi,\pi)$), and SAF (iii) State with DSC, one axial DDW ($\mathrm{DDW}_x$ 
or $\mathrm{DDW}_y$), and SAF (iv) State with DSC, $\mathrm{DDW}_x$, $\mathrm{DDW}_y$, and SAF. These four states exhaust all possible states relevant to answering the principal questions
in the recent developments on the observation of hidden charge order: whether the charge order is along the diagonal or along the axis of the Brillouin zone and whether 
it is of stripe or checkerboard pattern. Combinations which mix up axial and diagonal charge orders are not considered here simply because they break many more symmetries and experiments so far found no indication of such mixed states.
Comparing the energetics of states (i) and (ii), we can compare the relative energetics of states
with one diagonal 
charge order and two diagonal charge orders, perpendicular to each other. Similarly, comparing the energetics of states (iii) and (iv), 
we can then conclude the relative energetic competitiveness of the two types of axial charge orders; whether
it is uniaxial (stripe) or biaxial (checkerboard) charge order has lower energy. Comparing states (i) and (ii) versus states (iii) and (iv) on the other hand
gives the relative energetic competitiveness of axial versus diagonal charge orders, that is, determine whether diagonal or axial charge order has lower energy.
We will write the corresponding mean-field Hamiltonians for the states (i),(ii),(iii),(iv) from the action Eq. (\ref{HSeffaction}) in terms of appropriate spinor basis.

For state (i), the Hamiltonian contains $4\times 4$ matrix with 
basis $4\times 1$ spinor $\Psi^{\dag}_{\mathbf{k}}=(c^{\dag}_{\mathbf{k}\uparrow},c^{\dag}_{\mathbf{k}+\mathbf{Q}_{d+}\uparrow},c_{-\mathbf{k}\downarrow},c_{-(\mathbf{k}+\mathbf{Q}_{d+})\downarrow})$ 
where $\mathbf{Q}_d=\mathbf{Q}_{d+}=(\pi,\pi)$, as follows

\[
 H^{\mathrm{MF(i)}}_{\mathbf{k}}=
\]
\begin{equation}\label{MFHamiltI}
\left( \begin{array}{cccc}
\epsilon_k&\phi_+ f_k-M&\Delta^* f_k&0\\
\phi^*_+f_k-M^*&\epsilon_{k+Q_{d+}}&0&\Delta^*f_{k+Q_{d+}}\\
\Delta f_k&0&-\epsilon_{-k}&M^*-\phi_+ f_{k+Q_{d+}}\\
0&\Delta f_{k+Q_{d+}}&M-\phi^*_+f_{k+Q_{d+}}&-\epsilon_{-(k+Q_{d+})}
\end{array} \right)
\end{equation}
contributing to the full action Eq. (\ref{HSeffaction}) term of the form $\int_{\mathbf{k}}\Psi^{\dag}_{\mathbf{k}}H^{\mathrm{red}}_{\mathbf{k}}\Psi_{\mathbf{k}}$.
Despite the complicated expressions of those eigenvalues, in principle however, the Hamiltonian matrix in Eq. (\ref{MFHamiltI}) can be diagonalized analytically in closed-form 
for general case. Alternatively, one can diagonalize Eq. (\ref{MFHamiltI}) numerically, giving four eigenvalues and leading to new expression for the action Eq. (\ref{HSeffaction}) that can now be written as

\begin{equation}\label{HSeffaction0}
S_{[d^{\dag},d,\phi,\Delta]}=\int^{\beta}_0 d\tau[\frac{|\Delta|^2}{g_{DSC}}+\frac{|\phi_+|^2}{g_{DDW}}+\frac{|M|^2}{g_{SAF}}+\int_k \sum^4_{i=1}E^{i}_k {d^i}^{\dag}_{k}d^{i}_{k}]
\end{equation}
where $E^{i}_k,i=1,2,3,4$ are the four eigenvalues of Hamiltonian matrix Eq. (\ref{MFHamiltI}) and $d,d^{\dag}$ are the Bogoliubov quasiparticle operators. 
Integrating out the fermions and considering (imaginary) time-independent field theory (where the Hamiltonian only has no explicit dependence on imaginary time 
so that $\int^{\beta}_0 d\tau H = \beta H$), we obtain

\begin{equation}\label{HSeffHamiltonian1}
H_{[\Delta,\phi_+,M]}=\frac{|\Delta|^2}{g_{DSC}}+\frac{|\phi_+|^2}{g_{DDW}}+\frac{|M|^2}{g_{SAF}}-T\int_{k} \ln[\sum^4_{i=1}e^{-\frac{E^{i}_k}{T}}]
\end{equation} 
We also need to consider state with diagonal DDW along $\mathbf{Q}_{d-}=(-\pi,\pi)$ and compare its energy with the one with $\mathbf{Q}_{d+}=(\pi,\pi)$ as given above. 

For state (ii), the Hamiltonian contains $6\times 6$ matrix with 
basis $6\times 1$ spinor $\Psi^{\dag}_{\mathbf{k}}=(c^{\dag}_{\mathbf{k}\uparrow},c^{\dag}_{\mathbf{k}+\mathbf{Q}_{d+}\uparrow},c^{\dag}_{\mathbf{k}+\mathbf{Q}_{d-}\uparrow},c_{-\mathbf{k}\downarrow},c_{-(\mathbf{k}+\mathbf{Q}_{d+})\downarrow},c_{-(\mathbf{k}+\mathbf{Q}_{d-})\downarrow})$ 
where $Q_{d+}=(\pi,\pi),Q_{d-}=(-\pi,\pi)$, and the matrix is given by Eq. (\ref{MFHamiltII}) 
\comment{
\begin{equation}\label{MFHamiltII}
H^{\mathrm{MF(ii)}}_{\mathbf{k}}=\left( \begin{array}{cccccc}
\epsilon_k&\phi_+ f_k-M&\phi_-f_k-M&\Delta^* f_k&0&0\\
\phi^*_+f_k-M^*&\epsilon_{k+Q_{d+}}&0&0&\Delta^*f_{k+Q_{d+}}&0\\
\phi^*_-f_k-M^*&0&\epsilon_{k+Q_{d-}}&0&0&\Delta^*f_{k+Q_{d-}}\\
\Delta f_k&0&0&-\epsilon_{-k}&M^*-\phi_+ f_{k+Q_{d+}}&M^*-\phi_-f_{k+Q_{d-}}\\
0&\Delta f_{k+Q_{d+}}&0&M-\phi^*_+f_{k+Q_{d+}}&-\epsilon_{-(k+Q_{d+})}&0\\
0&0&\Delta f_{k+Q_{d-}}&M-\phi^*_-f_{k+Q_{d-}}&0&-\epsilon_{-(k+Q_{d-})}
\end{array} \right)
\end{equation}
}
Integrating out the fermions as before we obtain effective Hamiltonian
\begin{equation}\label{HSeffHamiltonian2}
H_{[\Delta,\phi_{\pm},M]}=\frac{|\Delta|^2}{g_{DSC}}+\frac{|\phi_+|^2}{g_{DDW}}+\frac{|\phi_-|^2}{g_{DDW}}+\frac{|M|^2}{g_{SAF}}-T\int_{k} \ln[\sum^6_{i=1}e^{-\frac{E^{i}_k}{T}}]
\end{equation}

For state (iii), the Hamiltonian contains $6\times 6$ matrix with 
basis $6\times 1$ spinor $\Psi^{\dag}_{\mathbf{k}}=(c^{\dag}_{\mathbf{k}\uparrow},c^{\dag}_{\mathbf{k}+\mathbf{Q}_X\uparrow},c^{\dag}_{\mathbf{k}+\mathbf{Q}_d\uparrow},c_{-\mathbf{k}\downarrow},c_{-(\mathbf{k}+\mathbf{Q}_X)\downarrow},c_{-(\mathbf{k}+\mathbf{Q}_d)\downarrow})$ 
where $Q_X=(\pi,0),Q_d=(\pi,\pi)$, and the matrix is given by Eq. (\ref{MFHamiltIII}) 
\comment{
\begin{equation}\label{MFHamiltIII}
H^{\mathrm{MF(iii)}}_{\mathbf{k}}=\left( \begin{array}{cccccc}
\epsilon_k&\phi_X f_k&-M&\Delta^* f_k&0&0\\
\phi^*_Xf_k&\epsilon_{k+Q_X}&0&0&\Delta^*f_{k+Q_X}&0\\
-M^*&0&\epsilon_{k+Q_d}&0&0&0\\
\Delta f_k&0&0&-\epsilon_{-k}&-\phi f_{k+Q_X}&M^*\\
0&\Delta f_{k+Q_X}&0&-\phi^*f_{k+Q_X}&-\epsilon_{-(k+Q_X)}&0\\
0&0&0&M&0&-\epsilon_{-(k+Q_d)}
\end{array} \right)
\end{equation}
}
We also need to consider the DSC+$\mathrm{DDW_y}$+SAF version of this state by simply replacing $Q_X$ with $Q_Y=(0,\pi)$.
The resulting effective Hamiltonian is
\begin{equation}\label{HSeffHamiltonian3}
H_{[\Delta,\phi_{X},M]}=\frac{|\Delta|^2}{g_{DSC}}+\frac{|\phi_X|^2}{g_{DDW}}+\frac{|M|^2}{g_{SAF}}-T\int_{k} \ln[\sum^6_{i=1}e^{-\frac{E^{i}_k}{T}}]
\end{equation} 

For state (iv), the Hamiltonian contains $8\times 8$ matrix with 
basis $8\times 1$ spinor $\Psi^{\dag}_{\mathbf{k}}=(c^{\dag}_{\mathbf{k}\uparrow},c^{\dag}_{\mathbf{k}+\mathbf{Q}_X\uparrow},c^{\dag}_{\mathbf{k}+\mathbf{Q}_Y\uparrow},c^{\dag}_{\mathbf{k}+\mathbf{Q}_d\uparrow}$,
$c_{-\mathbf{k}\downarrow},c_{-(\mathbf{k}+\mathbf{Q}_X)\downarrow},c_{-(\mathbf{k}+\mathbf{Q}_Y)\downarrow},c_{-(\mathbf{k}+\mathbf{Q}_d)\downarrow})$ where $Q_X=(\pi,0), Q_Y=(0,\pi), Q_d=(\pi,\pi)$, 
and the matrix is given by Eq. (\ref{MFHamiltIV}) 
\comment{
\begin{equation}\label{MFHamiltIV}
H^{\mathrm{MF(iv)}}_{\mathbf{k}}=\left( \begin{array}{cccccccc}
\epsilon_k&\phi_X f_k& \phi_Y f_k & -M&\Delta^* f_k&0&0&0\\
\phi^*_Xf_k&\epsilon_{k+Q_X}&0&0&0&\Delta^*f_{k+Q_X}&0&0\\
\phi^*_Yf_k&0&\epsilon_{k+Q_Y}&0&0&0&\Delta^*f_{k+Q_Y}&0\\
-M^*&0&0&\epsilon_{k+Q_d}&0&0&0&0\\
\Delta f_k&0&0&0&-\epsilon_{-k}&-\phi f_{k+Q_X}&-\phi f_{k+Q_Y}&M^*\\
0&\Delta f_{k+Q_X}&0&0&-\phi^*f_{k+Q_X}&-\epsilon_{-(k+Q_X)}&0&0\\
0&0&\Delta f_{k+Q_Y}&0&-\phi^*f_{k+Q_Y}&0&-\epsilon_{-(k+Q_Y)}&0\\
0&0&0&0&M&0&0&-\epsilon_{-(k+Q_d)}
\end{array} \right)
\end{equation}
}
with the resulting effective Hamiltonian 
\[
H_{[\Delta,\phi_{X(Y)},M]}=\frac{|\Delta|^2}{g_{DSC}}+\frac{|\phi_X|^2}{g_{DDW}}+\frac{|\phi_Y|^2}{g_{DDW}}+\frac{|M|^2}{g_{SAF}}
\]
\begin{equation}\label{HSeffHamiltonian4}
-T\int_{k} \ln[\sum^8_{i=1}e^{-\frac{E^{i}_k}{T}}]
\end{equation}

We are going to treat these effective Hamiltonians Eqs. (\ref{HSeffHamiltonian1}),(\ref{HSeffHamiltonian2}),(\ref{HSeffHamiltonian3}), and (\ref{HSeffHamiltonian4}) as the free energies (per unit cell) 
that we will minimize variationally \cite{NotePhases}.
Common approximation employed in studying cuprates with hot spots on Fermi surface is that in analyzing the fermion scattering process, only contributions
from fermions near hot spots are considered and the fermion kinetic energy dispersion $\epsilon_k$ is expanded to linear (or quadratic) order around the hot spots \cite{metlitsachdev}. 
In evaluating the effective Hamiltonians here however, rather than considering such expansion
in deriving the effective action for the order parameters, we use the full expression for energy dispersion $\epsilon_{k}$ but integrate over momenta within appropriate regime in Fourier space.

\section{Results}

We compute the energies given in Eqs.(\ref{HSeffHamiltonian1}),(\ref{HSeffHamiltonian2}),(\ref{HSeffHamiltonian3}), and (\ref{HSeffHamiltonian4}) and minimize them with respect to the corresponding 
variational mean field variables at given microscopic parameters ($t,t',t'',\mu,U,J,V_n,V_t,V_c$) as function of temperature in the low temperature ($T\lesssim 70 \mathrm{K}$) regime,
the relevant range of temperatures $T$ where DSC and DDW coexist, in addition to SAF. 
This calculation is aimed at determining the relative energetics of the four states described earlier, with emphasis
on the type of charge order characterizing each of those states,
which all we postulate to have $d$-wave character in their intra-unit cell structure $\Delta(\mathbf{k})$. 
Besides, the charge order is believed to exist below a critical temperature $T_{CO}$ 
all the way down to low temperatures to coexist with $d$-wave superconducting state that sets in at $T_c$ within certain range of hole doping levels.
In the phase diagram determined from experiments for compound $\mathrm{YBa_2Cu_3O_{6+\delta}}$ for example, charge order coexisting with superconductivity potentially occurs
within hole doping interval $0.10\lesssim p\lesssim 0.15$ where $50\lesssim T_c\lesssim 80 K$ \cite{Shekhter}.
The inclusion of DSC stresses the point that we consider charge order within superconducting background; the charge order coexists with superconductivity within superconducting dome.
The presence or absence of superconducting background is crucial and may change the relative energy balance between different orders in the four states considered above.

\begin{figure}
  \centering
 \includegraphics[scale=0.40]{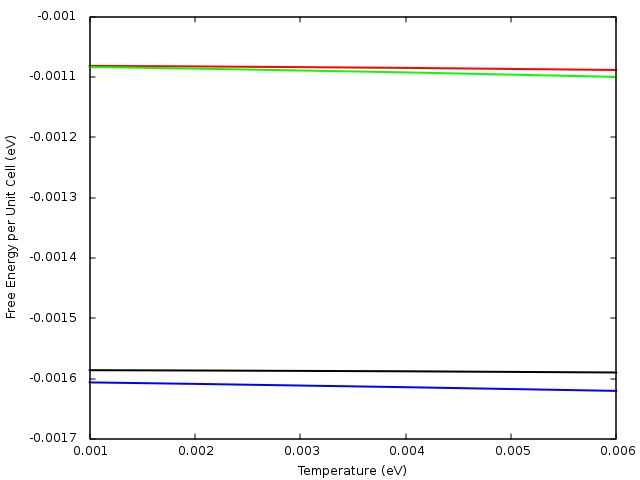}
  \caption{The variational energies $E_{var}=\langle H[\phi,\Delta]\rangle$ of states (i)(red: diagonal stripe),(ii)(green: diagonal checkerboard),(iii)(black: axial stripe),(iv)(blue: axial checkerboard) as functions of temperature $T$ in the low temperature $T\lesssim 70 \mathrm{K}$ regime in the commensurate limit.
 The parameters are $t=1.0,t'=0.3,\mu=0.0,V_n=0.5t, U=0.75 t, J=0.75t, V_c=0.8t,V_t=0$. The units of both axes are in eV. The solid lines are guide to the eye. }
\label{fig:EvsT}
\end{figure}

We eventually vary the microscopic parameters ($t,t',t'',\mu,U,J,V_n,V_t,V_c$) within appropriate interval which corresponds to the
regime of cuprate phase diagram we are interested in, which is the underdoped regime with the associated pseudogap phase. 
We actually find that, the conclusions to be described below, hold firmly all over this regime of the phase diagram and therefore do not
depend on a particular choice for a set of value of microscopic parameters used in the calculation.
It is to be noted that in the regime of phase diagram of interest here, the spin antiferromagnetism order parameter $M$ eventually vanishes, as has been found in various
previous theoretical studies, and we have $d$-wave superconducting state and orbital $d$-density wave orders remaining. 
With these considerations in mind, the result on relative energetic of the considered orders is presented in Fig.~\ref{fig:EvsT} where the units of both the temperature and free energy are in eV.

We can see that the two components of axial charge order with $(\pi,0)$ and $(0,\pi)$ are degenerate. So are the diagonal $(\pi,\pi)$ and $(-\pi,\pi)$ charge orders. 
This originates microscopically from the $x$ vs. $y$ symmetry of square lattice in the model Eq. (\ref{Laughlinmodel}).  
This is further supported by the finding that the effective Hamiltonian Eq. (\ref{HSeffHamiltonian1}) is invariant under $C_4$ rotations which implies exact degeneracy.
The degeneracy alone cannot decide on whether the two components of axial charge order and diagonal charge order should appear simultaneously or only one of each of them at once. 
However, the result of variational calculation shown in Fig.~\ref{fig:EvsT} clearly shows 
that the biaxial charge order, where both axial components of the charge order appear, has lower energy than uniaxial charge order, where only one of the two components appears. 
Similarly with the diagonal charge orders; bidiagonal charge order has lower energy than unidiagonal charge order.
The energy difference increases with temperature with the magnitude of the difference being slightly more significant between uniaxial and biaxial charge orders
than that between the unidiagonal and bidiagonal charge orders.
The energy difference is readily significant in the high temperature part in which charge order has been observed experimentally.
This finding therefore predicts that axial charge order in cuprates within superconducting dome, if it prevails, must be biaxial (checkerboard) in character rather than uniaxial (stripe).
Similar conclusion holds if the diagonal charge order prevails; it must be bidiagonal with both $Q_{d+}=(\pi,\pi)$ and $Q_{d-}=(-\pi,\pi)$ rather than only one of the two.
More importantly, we clearly see that the axial charge order consistently
has lower energy than the diagonal charge order and therefore axial charge order should be observed experimentally rather than diagonal order. 

\begin{figure}
  \centering
 \includegraphics[scale=0.40]{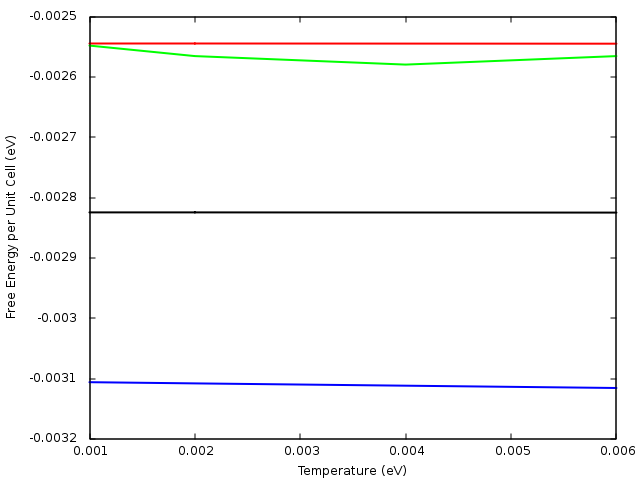}
  \caption{The variational energies $E_{var}=\langle H[\phi,\Delta]\rangle$ of states (i)(red: diagonal stripe),(ii)(green: diagonal checkerboard),(iii)(black: axial stripe),(iv)(blue: axial checkerboard) as functions of temperature $T$ in the low temperature $T\lesssim 70 \mathrm{K}$ regime in an incommensurate situation.
 The parameters are $t=1.0,t'=0.3,\mu=-0.30,V_n=0.5t, U=0.75 t, J=0.75t, V_c=0.8t,V_t=0$. The units of both axes are in eV. The solid lines are guide to the eye. }
\label{fig:EvsTincomm}
\end{figure}

The above are all the results in the commensurate limit. 
The treatment of general incommensurate case is very delicate. 
However, considering the absolute energetic competitiveness of axial charge order over diagonal one over large temperature range demonstrated in Fig.~\ref{fig:EvsT}, it is expected that
as the microscopic parameters are tuned adiabatically to give the more realistic incommensurate case, the conclusions above still hold very firmly. 
We check this prediction by considering cases slightly
away from commensurability by choosing chemical potential $\mu$ in the range $-0.5\lesssim \mu\lesssim 0$ with all other parameters the same as those used in Fig.~\ref{fig:EvsT}
with result shown in Fig.~\ref{fig:EvsTincomm}.
We observe that the axial charge order still has decisively lower energy than the diagonal one. 
Furthermore, the biaxial charge order also still has lower energy than uniaxial charge order and overall, biaxial charge order has the lowest variational free energy. 
In fact, the energetic advantage of the biaxial charge order over uniaxial one is even larger in this incommensurate case, 
compared to that of commensurate case. Real situations in experiments find density wave order with periodicity of $3\sim 4$ 
lattice spacings \cite{Hoffman}\cite{Ghiringelli}\cite{JChang}\cite{CominScience}, corresponding to $1.75\lesssim Q_0\lesssim 2.10$ and
can be reproduced by chemical potential in the range $-0.5\lesssim \mu \lesssim -0.3$ with all other parameters as used above \cite{Noteonparameters}. This realistic situation is thus well covered and 
this suggests that the energetic advantage of axial charge order concluded above from the commensurate limit should remain valid even to realistic situation.
This final result comes in remarkable agreement with experiments, as they indeed consistently observe axial charge order rather than diagonal one \cite{Hoffman,Vershinin,rcomin}.
Furthermore, our overall variational calculation predicts that within superconducting dome, the axial charge order should be biaxial (checkerboard) in character.

\section{Discussion}

We observe that the energetic preference for axial charge order over diagonal charge order arrives naturally in the mean field theory of Laughlin model Eq. (\ref{Laughlinmodel}) with DDW order scenario. 
While DDW order is assumed to begin with, no special emergent gauge symmetry is assumed in this work. The assumption of $d$-wave symmetry itself is directly responsible
for the energetic advantage of axial charge order over the diagonal one. 
The full effective
Hamiltonians Eqs. (\ref{HSeffHamiltonian1}-\ref{HSeffHamiltonian4}) can be expanded in terms of order parameters $\Delta$, $\phi$, and $M$. 
The expansion automatically includes higher than quadratic (cubic, quartic, \ldots) order terms 
which describe the interplay between the three orders we are considering. The energetic calculations presented above clearly suggests that the delicate interplay unambiguously 
singles out and prefers axial charge order over diagonal charge order.
The interplay between DDW charge order and superconductivity is embodied naturally in the effective
Hamiltonians Eqs. (\ref{HSeffHamiltonian1}-\ref{HSeffHamiltonian4}) obtained upon integrating out the fermions including 
the feedback effect of superconductivity to charge order, which suppresses 
diagonal charge order and favors axial charge order.
It is to be noted that the charge orders we consider here have superconducting state in the background; that is, 
they all occur within superconducting dome. 
It is interesting to note that within superconducting dome, biaxial charge order (checkerboard) is found here to be the most favorable state energetically, while for charge order outside superconducting dome,
it was found that uniaxial (stripe) charge order prevails \cite{chubukov}.

Some recent experiments have not been able to conclusively decide on the uniaxiality versus biaxiality of the charge order \cite{rcomin}. 
However, other more recent experiments \cite{inelasticxray} from x-ray scattering apparently provide strong evidence that biaxial state really prevails 
as scattering peaks are observed at wave vectors $(Q_0,0)$ and $(0,Q_0)$ simultaneously. Ultrasound measurement also suggests that the charge order is biaxial \cite{LeBoeuf}.
Theoretically, the exact degeneracy between the two orthogonal components of axial charge order provides the first hint that they could possibly appear simultaneously.
This is especially the case when there is no perturbation which tips off the balance between the tendencies to favor one of the two orthogonal components.
In this work, complete $C_4$ ($x$ vs. $y$) symmetry is assumed within DDW scenario and it should come as no surprise that the mean field effective hamiltonian gives rise to the expected
exact degeneracy. Further consideration that may enhance the possibility for simultaneous appearance is the absence of relatively large potential barrier ('wall') that separates the two orthogonal states.
When such condition is satisfied, the overlap between the two states is significant and the system can relatively easily resonate between the two orthogonal states rather than being trapped in only 
one of the two 'wells'.

In this work, our result from straightforward variational mean field study of fully microscopic model with DDW scenario clearly suggests that the axial charge order must indeed  be biaxial (checkerboard)
due to energetic competitiveness of the biaxial charge order, as shown in Figs.~(\ref{fig:EvsT}) and ~(\ref{fig:EvsTincomm}).
The results have shown agreement with experiments on the energetic preference for axial charge order and it remains to be seen whether concensus on the predicted checkerboard nature of the axial
charge order will be achieved. It is interesting to see whether variational mean field calculation on other microscopic models, such as the 3-band model of cuprates would
also deliver the same conclusions, as studied recently \cite{Bulut}\cite{ThomsonSachdev}.

The question of whether the charge density wave in cuprates is stripe or checkerboard is indeed a long standing issue. Some researchers \cite{BergNJPstripedSC} have proposed phenomenological theory in favor of stripe order as the universal character of the superconducting state and the ordered state in the pseudogap regime. In practice and experiment, this problem is indeed difficult to resolve.  The real-space pattern of the charge density wave is often very sensitive to disorder \cite{Robertson}\cite{Maestro}. What is measured in experiment such as by x-ray scattering is normally the momentum space profile of the charge density wave, in the form of structure factor.  Yet another x-ray scattering experiment \cite{CominStripe} on YBCO compound recently observed the presence of peaks in the structure factor both for momenta $\mathbf{Q}_x=(\pm Q_0,0)$ and
$\mathbf{Q}_y=(0,\pm Q_0)$ at temperature close to but slightly above the superconducting critical temperature $T_c$, where there is no superconductivity. In our view, if the physics behind the result came entirely from the 2d physics of cuprates and there were only a single domain of ordered state on each layer of cuprates, then it should immediately imply checkerboard order. In that work, the treatment assumes two-dimensional physics of the plane of the cuprates with multi-domains of charge ordered patches on each CuO$_2$ plane and the structure factor was further analyzed in terms of its symmetries in which they observed anisotropy between the structure factor peaks at $\mathbf{Q}_x$  and $\mathbf{Q}_y$.Rather than concluding it as signature of checkerboard order with a single domain extending over each plane of cuprates, it was argued that the result indicates 90 degrees rotated regions of $\mathbf{Q}_x$  and $\mathbf{Q}_y$  charge stripe domains in each CuO$_2$ plane layer.However, experimentally, the structure factor measured from x-ray scattering probes the signal scattered by all the layers of the cuprates sample.As such, the signal probes the 3d physics of the cuprate material despite the poor coherence of the charge density wave across the CuO$_2$ planes. In addition, even though the structure factor is measured as a function of planar wavevectors $S(Q_x,Q_y)$, the result may actually represent the contribution of many CuO$_2$ planes, because one cannot isolate just a single CuO$_2$ plane.Furthermore, the problem with multi-domain scenario is that there is no definite knowledge of how the domains should be distributed; what are the sizes and the orientations as well as the positioning of the domains.The well-defined and patterned peaks in the structure factor as observed in the x-ray scattering can only occur when there is a strong coherence in the distribution of the domains, the experimental evidence for which is not established.Therefore, an alternative interpretation with stripy charge ordering is that the peaks imply a crisscrossed stripe order, where the stripe order alternates with wavevector $\mathbf{Q}_x$ and $\mathbf{Q}_y$ between adjacent layers and extends over the full area of each CuO$_2$ plane as a single domain \cite{crisscross}. This is plausibly indeed the true scenario for charge density wave outside the superconducting regime. 

Existing microscopic theoretical works based on spin-fermion model \cite{chubukov} with linear hot spot approximation for computing effective free energy for charge density wave without superconductivity found result which also favors stripe order. However, at the same time they made a precautionary remark that in the presence of superconductivity, there will be a slight change in the calculation due to the infrared cutoff given by the superconducting gap in the momentum integrals of the Feynman diagram  which however can give result which favors checkerboard order. The model they considered is 2d model of cuprates as considered in our work. The result of our work basically implies that this latter situation as they cautioned is exactly what happens when one goes beyond linear hot spot approximation and considers superconductivity in coexistence with charge density wave; the checkerboard order prevails. Our result is also in nice agreement with recent work \cite{wangchubukovstripevschecker} who used free energy calculation computed perturbatively from spin-fermion model and found that over practically 
all temperature ranges, bidirectional (checkerboard) charge order prevails. It is only deep within the charge ordered phase at lower doping level side outside the superconducting dome that the stripe charge density wave can take over the control as the ground state.This latter regime is beyond the scope of our work.       

In general, superconductivity and charge density wave are two competing orders; they compete for the same hot spot fermions in order to prevail as the ground state. The presence of superconductivity thus in general provides feedback which tends to suppress the charge order.  The charge order itself has several different possible configurations in terms of the wavevector $\mathbf{Q}$ and the structure factor $\Delta^{\mathbf{Q}}_{\mathbf{k}}$. Within the SU(2) symmetric theory of linearized hot spot approximation \cite{metlitsachdev}, the  d-wave superconductivity and diagonal charge density wave are degenerate leading instabilities, while axial charge order is subleading (together with pair density wave). Following works \cite{chowdurysachdev} then found that the diagonal charge order is suppressed more strongly than the axial charge order by the feedback from superconductivity so that axial charge order may eventually prevail, at least within a window in parameter space. Again, stripe order  along axial direction was found within linear hot spot approximation. There is however a precaution that upon including the curvature part of the dispersion around the hot spot, the conclusion of linear hot spot approximation is valid only above a certain temperature scale and may cease below that scale. In the temperature regime below this scale therefore, checkerboard order may actually prevail, which corresponds to the checkerboard order concluded in our work.

The x-ray scattering result \cite{CominStripe} shows anisotropy at each of the four peaks in the structure factor $S(Q_x,Q_y)$. The questions then are what the sources of this anisotropy are and what its consequences are. Does the anisotropy determine the energetic of the system or simply lead to minor asymmetry in the charge order, manifesting as anisotropic peaks in the structure factor?Does anisotropy immediately imply stripe order which breaks $C_4$ symmetry of the lattice?
One of the possible sources of anisotropy is the orthorhombicity of the crystal of cuprate material in which the lattice spacing $a$ is different from the lattice spacing $b$. In some cuprate materials such as YBCO, it is found that $a<b$. However, if the orthorhombicity were to give rise to stripe charge density wave, then naively and very intuitively, one should have stripe order in just one of the two directions only, but not both, because we can only have $a<b$ and not $a>b$ or vice versa for a given material. This reasoning therefore suggests that orthorhombicity does not explain the crisscrossed stripe order as the origin of the observed simultaneous peaks at $\mathbf{Q}_x$ and $\mathbf{Q}_y$. 
In fact, orthorhombicity might contribute to the anisotropy of the orthogonal peaks observed in x-ray scattering, assuming that the peaks originate from a checkerboard order.
If the physics is indeed entirely two-dimensional with single domain of ordered state extending over the whole CuO$_2$ plane, then the orthogonal peaks observed in x-ray scattering can only be explained by checkerboard order. Crisscrossed stripe order is likely only if the physics is effectively three-dimensional, for which stripe order in each layer can be the prevailing ground state, which would be a consistent scenario to explain x-ray scattering experimental result. 

Another possible source of anisotropy is the presence of CuO chain, known to appear in YBCO in appropriate doping regime. It is believed that CuO chain serves as a reservoir of charge for the transport that mainly occurs on the CuO$_2$ plane. As such, it is indeed physically intuitive that the CuO chain, being directed parallel to the b axis, gives rise to anisotropy in the hopping integral, which characterizes the metallic property of the system. 
In our perspective, the presence of CuO chain simply contributes to an anisotropy in the precise quantitative detail of the charge density wave, such as the relative strength of the modulations of charge density along $x$ and $y$ or the shape of structure factor peaks at $\mathbf{Q}_x$ and $\mathbf{Q}_y$, but is not the one which decides the qualitative character of the charge order, that is, whether one has checkerboard order or crisscrossed stripe order, assuming a single domain on each CuO$_2$ plane.In fact, we believe that the anisotropy of the peaks in the structure factor found in x-ray scattering is contributed by the presence of CuO chain.If the CuO chain were the origin of stripe order, since the chain is directed only along b axis and not along a axis, the stripe order due to the CuO chain would be along one direction, rather than alternating along a and b directions between adjacent CuO$_2$ planes, in the crisscrossed stripe scenario that may explain the orthogonal peaks in the structure factor observed in x-ray scattering experiment on YBCO at slightly above the superconducting critical temperature.Similarly, at temperatures below $T_c$ where the charge order coexists with superconductivity and we have checkerboard order, the CuO chain merely gives rise to quantitative anisotropy in the strength of charge modulations in the x and y directions.The stability of the checkerboard order against anisotropy in the kinetic hopping integral $t$ due to CuO chain is guaranteed by the fact that the energetic of the state is determined more by the interaction terms in the Hamiltonian rather than the kinetic terms.This is evident because it is the interaction terms that give rise to the spin fluctuations that mediate the scattering of fermions between the hot spots and give rise to the significant difference in the energetics of the four different charge ordered states considered in this work.

The above analysis suggests that one can have transition from stripe to checkerboard order if there is a change in the effective dimensionality of the physics of the system. We can say that the charge density wave is crisscrossed stripe when the system is effectively 3d and checkerboard when it is effectively 2d. The possible driving force for the transition between the two cases is interlayer coupling. The above analysis suggests that at high temperature with no superconductivity, interlayer coupling may be significant that the system is effectively 3d and one has crisscrossed stripe as the ground state, which gives rise to simultaneous peaks in the structure factor at both $\mathbf{Q}_x$ and $\mathbf{Q}_y$. At lower temperatures, in the presence of superconductivity, interlayer coupling is negligible that the physics is entirely 2d and we have checkerboard order that also gives rise to simultaneous peaks at the two orthogonal wavevectors. This argument also explains why in our calculation, we always obtain checkerboard order as the lowest energy ground state: Our theory is defined on purely 2d system for which the two orthogonal directions are on equal footing and therefore give rise to checkerboard order, which extends over the whole CuO$_2$ plane.  Our free energy calculation also includes all order in the expansion of energy dispersion around hot spot, rather than just to linear order, and this inclusion of all orders secures the checkerboard order to be the prevailing ground state. Existing theoretical works obtain stripe order from purely 2d model because they retained only linear order term in the expansion of fermion dispersion around hot spot and without considering superconductivity in coexistence with charge density wave. This existing result indeed seems to explain the observed anisotropic orthogonal peaks in x-ray scattering, if the stripe is crisscrossed between adjacent copper planes. 

Remarkably, such dimensional crossover from effective three-dimensional physics at high temperatures to two-dimensional physics at lower temperatures is indeed what has been observed in the available experimental reports in the literature, from the 90's \cite{dimencross1}\cite{dimencross2}. There it was found that at high temperatures, YBCO behaves as anisotropic 3d system with strong interlayer tunneling whereas at lower temperatures, it behaves as weakly-coupled 2d system with Josephson coupling. This provides support to our physical hypothesis above regarding the relation between effective dimensionality of the cuprate system and the directionality of charge order: At high temperatures, the charge order is possibly crisscrossed stripe order, made possible by strong interlayer tunneling and the effective three-dimensionality of the system.At lower temperatures, the charge order is checkerboard order, as this is the lowest energy configuration in effective two dimensional system.In both cases, the resulting structure factor contains peaks at the two orthogonal wavevectors, as seen in the x-ray scattering experiment, and we have assumed a single domain of charge ordered state extending over the whole area of each CuO$_2$ plane. This scenario would represent an interesting novel proposal to answer the question of the directionality of the charge order in cuprates. While this issue has not been fully resolved at the moment, our work helps shed light on the energetics of the different configurations of charge ordered state in underdoped cuprates and on the physics of the pseudogap regime in broader perspective.

\begin{acknowledgment}


I would like to thank D. Poilblanc, R. Ramazashvili, C. Proust, D. LeBoeuf, and F. Laliberte for very useful discussions. 
This study was supported by the Grant n$^\mathrm{o}$ ANR-10-LABX-0037
in the framework of the «Programme des Investissements d'Avenir» of France. 

\end{acknowledgment}


\appendix

\onecolumn

\section{Detailed Expressions of the Effective Hamiltonians for Different States}

We give here the detailed expressions of the Hamiltonians describing the different states discussed in the main text.
For the state (ii), the Hamiltonian contains $6\times 6$ matrix with 
basis $6\times 1$ spinor $\Psi^{\dag}_{\mathbf{k}}=(c^{\dag}_{\mathbf{k}\uparrow},c^{\dag}_{\mathbf{k}+\mathbf{Q}_{d+}\uparrow},c^{\dag}_{\mathbf{k}+\mathbf{Q}_{d-}\uparrow},c_{-\mathbf{k}\downarrow},c_{-(\mathbf{k}+\mathbf{Q}_{d+})\downarrow},c_{-(\mathbf{k}+\mathbf{Q}_{d-})\downarrow})$ 
where $Q_{d+}=(\pi,\pi),Q_{d-}=(-\pi,\pi)$, and the matrix is given by

\begin{equation}\label{MFHamiltII}
H^{\mathrm{MF(ii)}}_{\mathbf{k}}=\left( \begin{array}{cccccc}
\epsilon_k&\phi_+ f_k-M&\phi_-f_k-M&\Delta^* f_k&0&0\\
\phi^*_+f_k-M^*&\epsilon_{k+Q_{d+}}&0&0&\Delta^*f_{k+Q_{d+}}&0\\
\phi^*_-f_k-M^*&0&\epsilon_{k+Q_{d-}}&0&0&\Delta^*f_{k+Q_{d-}}\\
\Delta f_k&0&0&-\epsilon_{-k}&M^*-\phi_+ f_{k+Q_{d+}}&M^*-\phi_-f_{k+Q_{d-}}\\
0&\Delta f_{k+Q_{d+}}&0&M-\phi^*_+f_{k+Q_{d+}}&-\epsilon_{-(k+Q_{d+})}&0\\
0&0&\Delta f_{k+Q_{d-}}&M-\phi^*_-f_{k+Q_{d-}}&0&-\epsilon_{-(k+Q_{d-})}
\end{array} \right)
\end{equation}

For the state (iii), the Hamiltonian contains $6\times 6$ matrix with 
basis $6\times 1$ spinor $\Psi^{\dag}_{\mathbf{k}}=(c^{\dag}_{\mathbf{k}\uparrow},c^{\dag}_{\mathbf{k}+\mathbf{Q}_X\uparrow},c^{\dag}_{\mathbf{k}+\mathbf{Q}_d\uparrow},c_{-\mathbf{k}\downarrow},c_{-(\mathbf{k}+\mathbf{Q}_X)\downarrow},c_{-(\mathbf{k}+\mathbf{Q}_d)\downarrow})$ 
where $Q_X=(\pi,0),Q_d=(\pi,\pi)$, and the matrix is given by

\begin{equation}\label{MFHamiltIII}
H^{\mathrm{MF(iii)}}_{\mathbf{k}}=\left( \begin{array}{cccccc}
\epsilon_k&\phi_X f_k&-M&\Delta^* f_k&0&0\\
\phi^*_Xf_k&\epsilon_{k+Q_X}&0&0&\Delta^*f_{k+Q_X}&0\\
-M^*&0&\epsilon_{k+Q_d}&0&0&0\\
\Delta f_k&0&0&-\epsilon_{-k}&-\phi f_{k+Q_X}&M^*\\
0&\Delta f_{k+Q_X}&0&-\phi^*f_{k+Q_X}&-\epsilon_{-(k+Q_X)}&0\\
0&0&0&M&0&-\epsilon_{-(k+Q_d)}
\end{array} \right)
\end{equation}

For the state (iv), the Hamiltonian contains $8\times 8$ matrix with 
basis $8\times 1$ spinor $\Psi^{\dag}_{\mathbf{k}}=(c^{\dag}_{\mathbf{k}\uparrow},c^{\dag}_{\mathbf{k}+\mathbf{Q}_X\uparrow},c^{\dag}_{\mathbf{k}+\mathbf{Q}_Y\uparrow},c^{\dag}_{\mathbf{k}+\mathbf{Q}_d\uparrow}$,
$c_{-\mathbf{k}\downarrow},c_{-(\mathbf{k}+\mathbf{Q}_X)\downarrow},c_{-(\mathbf{k}+\mathbf{Q}_Y)\downarrow},c_{-(\mathbf{k}+\mathbf{Q}_d)\downarrow})$ where $Q_X=(\pi,0), Q_Y=(0,\pi), Q_d=(\pi,\pi)$, 
and the matrix is given by 

\begin{equation}\label{MFHamiltIV}
H^{\mathrm{MF(iv)}}_{\mathbf{k}}=\left( \begin{array}{cccccccc}
\epsilon_k&\phi_X f_k& \phi_Y f_k & -M&\Delta^* f_k&0&0&0\\
\phi^*_Xf_k&\epsilon_{k+Q_X}&0&0&0&\Delta^*f_{k+Q_X}&0&0\\
\phi^*_Yf_k&0&\epsilon_{k+Q_Y}&0&0&0&\Delta^*f_{k+Q_Y}&0\\
-M^*&0&0&\epsilon_{k+Q_d}&0&0&0&0\\
\Delta f_k&0&0&0&-\epsilon_{-k}&-\phi f_{k+Q_X}&-\phi f_{k+Q_Y}&M^*\\
0&\Delta f_{k+Q_X}&0&0&-\phi^*f_{k+Q_X}&-\epsilon_{-(k+Q_X)}&0&0\\
0&0&\Delta f_{k+Q_Y}&0&-\phi^*f_{k+Q_Y}&0&-\epsilon_{-(k+Q_Y)}&0\\
0&0&0&0&M&0&0&-\epsilon_{-(k+Q_d)}
\end{array} \right)
\end{equation}


\begin{thebibliography}{9}
\bibitem{GScorrection}The (commensurate) DDW order parameter will eventually involve only
the imaginary part of the full bond order parameter $\langle c^{\dag}_{i\sigma}c_{j\sigma}\rangle$: $\phi$ is to be taken as purely imaginary with the phase fixed to $\pi/2$ \cite{C-L-M-N}.
That is, $\phi=|\phi|\exp(i\pi/2)=i|\phi|$.
\bibitem{NotePhases} The order parameters are taken to be complex-valued in general. 
With $d_{x^2-y^2}$ $d$-wave SC state assumed, $\Delta$ eventually takes real value. The $M$ physically represents magnetization and is therefore real-valued. 
As noted earlier \cite{GScorrection}, since $\phi$ is purely imaginary, then only the magnitudes of $\Delta$, $\phi$, and $M$ 
are varied and the phases drop out as they are fixed to $0$ for DSC and SAF and $\pi/2$ for DDW respectively.
\bibitem{Noteonparameters}The conclusions derived from this calculation are independent of parameters.
A proper set of parameters has been used in the calculation as an example to illustrate the physics that applies in the whole underdoped pseudogap regime
of the cuprate phase diagram under consideration.
\end{thebibliography}
\end{document}